\documentclass[a4paper,11pt]{article}
\pdfoutput=1
\usepackage{jheppub}
\usepackage{graphicx}
\usepackage{amsmath}
\usepackage{amsfonts}
\usepackage{amssymb}
\usepackage[colorlinks=true,linkcolor=blue]{hyperref}
\usepackage{bm}
\usepackage{empheq}
\usepackage[super]{nth}

\def\half{\frac{1}{2}}
\def\OO{\mathcal{O}}
\def\LL{\mathcal{L}}
\def\MM{\mathcal{M}}
\def\VV{\mathbf{V}}
\def\be{\begin{equation}}
\def\ee{\end{equation}}
\def\Eq#1{Eq.~\eqref{#1}}

\def\epspot{\varepsilon_{\mathrm{pot}}}
\def\epskin{\varepsilon_{\mathrm{kin}}}
\def\epsgrad{\varepsilon_{\mathrm{grad}}}

\title{Hydrodynamics as $v_s \to c$}
\author{Guy D.\ Moore}

\affiliation{Institut f\"ur Kernphysik, Technische Universit\"at Darmstadt\\
Schlossgartenstra{\ss}e 2, D-64289 Darmstadt, Germany}
\emailAdd{guy.moore@physik.tu-darmstadt.de}

\abstract{
I present the simplest 3+1 dimensional quantum field theory for which
the speed of sound can be arbitrarily close to the speed of light.
Examining the hydrodynamics, I find cases where the shear viscosity is
finite, but the ``shear relaxation coefficient'' appears always to be
divergently large.
}

\keywords{relativistic hydrodynamics, equation of state}
\begin{document}
\maketitle
\section{Introduction}
\label{sec:intro}

It appears that the high-density equation of state of QCD at large
baryon chemical potential, such as exists in neutron star interiors,
requires a startlingly high speed of sound to support the largest-mass
neutron stars so far observed
\cite{Koehn:2024set}.
This has raised the general theoretical question -- how high can the
speed of sound be in physically reasonable quantum field theories?
It has been proposed
\cite{Cherman:2009tw}
that the speed of sound $v_s$ in theories \textsl{without} chemical
potentials should not exceed the conformal value%
\footnote{%
I will use natural units $\hbar=1$, $c=1$ and the mostly-positive
spacetime metric $\eta_{\mu\nu} = \mathrm{Diag}[-1,1,1,1]$.}
$v_s^2 = 1/3$.  (We will mention a counterexample in the discussion,
for which $v_s^2$ mildly exceeds $1/3$.)

Recently, Hippert Noronha and Romatschke have argued that there should
be an upper bound on the speed of sound,
$v_s^2 < 0.781$, based on hydrodynamical arguments
\cite{Hippert:2024hum}.
Their argument assumes that the theory in question has ``ordinary''
hydrodynamical behavior -- in particular, the shear viscosity
relaxation rate $\tau_\pi$ should exist.  On the other hand, in the
presence of chemical potentials, there is a known example where the
speed of sound can be arbitrarily close to the speed of light.  Son
and Stephanov \cite{Son:2000by} have shown that, within chiral
perturbation theory, QCD at nonzero isospin chemical potential obeys
$P/\varepsilon = (\mu_I^2 - m_\pi^2)/(\mu_I^2 + 3 m_\pi^2)$,
which in the chiral limit $m_\pi^2 \to 0$ returns
$P/\varepsilon \to 1$ up to corrections suppressed by
$m_\pi^2/\mu_I^2$ and $\mu_I^2 / (4\pi f_\pi)^2$ -- the latter arising
from the limitations of chiral perturbation theory.
Therefore the speed of sound approaches 1 in this theory in the regime
$m_\pi^2 \ll \mu^2 \ll (4\pi f_\pi)^2$.
The physical origin and meaning of this counterexample
is not immediately obvious.  Nor is it clear how it evades the bound
proposed in Ref.~\cite{Hippert:2024hum}.

The purpose of this brief note is to present the simplest example of a
quantum field theoretical system where $v_s^2$ can approach 1,  to
clearly explain the physical origin of this effect, and to study the
hydrodynamics of such theories.
The simplest theory we can find where $v_s^2 \to 1$ is $N$-component
scalar $\lambda \varphi^4$ theory with $N \geq 2$,
with $O(N)$ symmetry and a potential
which leads to the spontaneous breaking of the $O(N)$ symmetry.
This theory has $N(N-1)/2$ Noether currents with associated charges.
In the case with a chemical potential $\mu$ for one of these charges
satisfying $\mu \sim T \ll m_r$ with $m_r$ the heavy mass of radial
scalar excitations, we show that the speed of sound approaches 1.
For $N=2$ this theory has an exponentially large shear viscosity,
but for $N>2$ the shear viscosity proves to be finite.
However, $\tau_\pi$, defined in terms of a Kubo relation,
proves to diverge in either case, which explains why the
arguments of Hippert \textsl{et al} do not apply.

The next section demonstrates why $O(N)$ scalar field theory can
support a speed of sound arbitrarily close to the speed of light.
Section \ref{sec:hydro} investigates the hydrodynamics of this model,
and shows that the shear viscosity is finite for $N>2$ but the shear
relaxation rate diverges.
The paper ends with a brief discussion.

\section{Examining the O(N) model with a large chemical potential}
\label{sec:ON}

Consider $N$-component scalar field theory with scalar fields
$\varphi_a$, $a = 1,\ldots,N$ with the most general $O(N)$ respecting,
renormalizable Lagrangian
\begin{equation}
  \label{Lagrangian}
  - \LL = \half \partial_\mu \varphi_a \partial^\mu \varphi_a
  - \frac{m_r^2}{4} \varphi_a \varphi_a
  + \frac{\lambda}{8} \left( \varphi_a \varphi_a \right)^2 \,,
\end{equation}
where repeated indices are summed and
where we have written the $-m_r^2/4$ term with a minus sign so that
the potential is symmetry breaking and $m_r$ will represent the mass
of the ``radial'' excitation about the symmetry-broken minimum.
Defining the vacuum expectation value $v^2 = m_r^2/\lambda$,
we can rewrite the potential terms (after shifting by a constant) as
\begin{equation}
  V(\varphi_a \varphi_a) = \frac{\lambda}{8}
  \left( \varphi_a \varphi_a - v^2 \right)^2 \,.
\end{equation}
The associated energy density $\varepsilon = T^{00}$ is
\begin{equation}
  \label{Hamiltonian}
  \varepsilon =
  \half \partial_t \varphi_a \partial_t
  \varphi_a + \half \partial_i \varphi_a \partial_i \varphi_a
  + V(\varphi_a \varphi_a)  \,.
\end{equation}
Here as usual $i$ runs over the spatial directions and repeated
indices are summed.
We will assume that $\lambda \ll 1$ and that the renormalization scale
and scheme have been chosen such that the renormalizations of $v$ and
$\lambda$ are taken into account.
The classical vacuum state is that one scalar has a vacuum value equal to $v$
and all others are zero.  Without loss of generality we can pick the
$\varphi_1$ direction for the vacuum value, in which case $\varphi_1$
excitations have a mass $m_r$, which we refer to as the radial mass%
\footnote{%
Some would call it a Higgs mass, but properly speaking this should be
reserved for such a radial excitation when the symmetry is gauged.},
and the other $\varphi_{2\ldots N}$ directions are massless Goldstone
modes.  We will primarily be interested in the case where both the
temperature and any chemical potentials are small compared to $m_r$.

We can similarly define the pressure, which is
\begin{equation}
  \label{Pressure}
  P = \frac{T^{ii}}{3}
  =  \half \partial_t \varphi_a \partial_t \varphi_a
  - \frac{1}{6} \partial_i \varphi_a \partial_i \varphi_a
  - V(\varphi_a \varphi_a) \, .
\end{equation}
The same three terms appear, but with different coefficients:
\begin{enumerate}
\item
  Potential energy $\epspot = (\lambda/8)(\varphi_a \varphi_a-v^2)^2$
  contributes oppositely to the energy density $\varepsilon$ and
  to the pressure $P$:  $P_{\mathrm{pot}} = - \epspot$.
  This is familiar from cosmology as the behavior of a cosmological
  constant, which can also arise from a scalar field potential and
  plays a central role in inflation \cite{Guth:1980zm}.
\item
  Gradient energy
  $\epsgrad = \half \partial_i \varphi_a \partial_i \varphi_a$
  satisfies $P_{\mathrm{grad}} = -\epsgrad/3$.
\item
  Kinetic energy
  $\epskin = \half \partial_t \varphi_a \partial_t \varphi_a$
  contributes equally to $\varepsilon$ and $P$:
  $P_{\mathrm{kin}} = \epskin$.
  This is familiar in cosmology from the idea of \textsl{kination}
  \cite{Joyce:1996cp,Spokoiny:1993kt}.
\end{enumerate}
A large speed of sound $v_s^2 \equiv dP/d\varepsilon \simeq 1$
requires $P \simeq \varepsilon$ and therefore a domination of the
energy density by $\dot\varphi$ terms, $\epskin \gg \epspot,\epsgrad$.
This can occur if the energy
is dominated by the spatially homogeneous motion of the scalar field
through field-space, provided that this scalar condensate does not
explore large values of the potential energy.
This will occur if the scalar field revolves around the vacuum
manifold, which will occur in the case of a chemical potential $\mu$
for a conserved Noether current in the case $\mu^2 \ll m_r^2$, as we
see next.

There are $N(N-1)/2$ conserved Noether currents
\begin{equation}
  \label{Noether1}
  j^\mu_{ab} \equiv -\varphi_a \partial^\mu \varphi_b
  + \varphi_b \partial^\mu \varphi_a \,,
  \qquad
  a < b .
\end{equation}
The associated charges are $Q_{ab} = \int d^3 x j^0_{ab}$.
Suppose we apply
a chemical potential $\mu$ for
$Q_{12} = \int d^3 x \left( \varphi_1 \partial_t \varphi_2
- \varphi_2 \partial_t \varphi_1 \right)$.
This is like an ``angular momentum'' in the $(\varphi_1,\varphi_2)$ plane
which will cause the fields to rotate around in a circle within this
plane:
\begin{equation}
  \label{background}
  ( \, \varphi_1(t) \, , \, \varphi_2(t) \, ) =
  \left(\, v_\mu \cos(\omega t)\, , \, v_\mu \sin(\omega t)\, \right).
\end{equation}
The mean value of the field need not equal its vacuum value so we have
written it as $v_\mu$.
The frequency of rotation $\omega$ and condensate value $v_\mu$ are
solved by minimizing $\int d^3 x \left(\varepsilon - \mu j^0 \right)$:
\begin{align}
  \label{HminusQ}
  \varepsilon - \mu j^0 & = \half \omega^2 v_\mu^2 - \mu \omega v_\mu^2
  + \frac{\lambda}{8}(v_\mu^2 - v^2)^2
  \\
  \label{fixomega}
  0 = \frac{d(\epsilon - \mu j^0)}{d\mu} & = (\omega - \mu) v_\mu^2
  & \omega & = \mu
  \\
  \label{fixvev}
  0 = \frac{d(\epsilon - \mu j^0)}{dv^2_\mu} & = - \half \mu^2
  + \frac{\lambda}{4} (v_\mu^2 - v^2)
  & v_\mu^2 & = v^2 + \frac{2\mu^2}{\lambda} \,.
\end{align}
The frequency of rotation is precisely the chemical potential.
The increase in the vacuum expectation value can be understood as
follows.  The scalar field undergoes a field-space acceleration of
$\omega^2 v_\mu$, which must be compensated by a radial restoring
force:
\begin{equation}
  \label{explainvev}
  \mu^2 v_\mu = \frac{dV}{dv_\mu} = \frac{\lambda}{2} v_\mu (v_\mu^2-v^2)
  \quad \Rightarrow \quad
  v_\mu^2 = v^2 + \frac{2\mu^2}{\lambda} \,.
\end{equation}

\begin{figure}[tbh]
  \centerline{
    \includegraphics[width=0.8\textwidth]{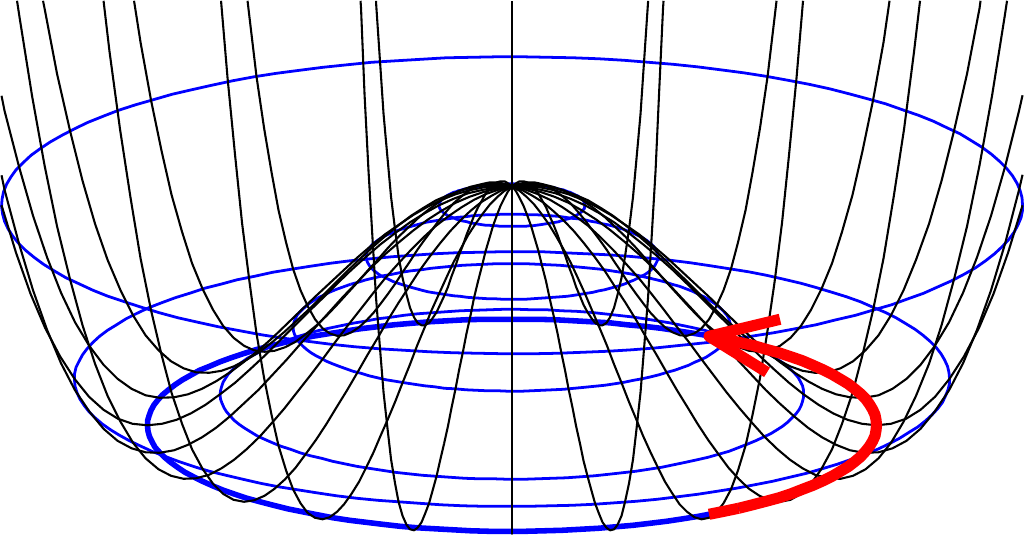}}
  \caption{\label{fig:cartoon}
    Cartoon of the effective potential and the path the scalar field
    takes around the valley in the potential in the presence of a
    chemical potential.}
\end{figure}

The resulting kinetic and potential energies are respectively
$\epskin = \mu^2 v_\mu^2/2 = \mu^2 v^2/2 + \mu^4/\lambda$ and
$\epspot = \mu^4 / 2\lambda$.
In the limit $\lambda v^2 \gg \mu^2$ we have $\epskin \gg \epspot$.
Note that $\lambda v^2 = m_r^2$, so this limit is the same as
$\mu^2 \ll m_r^2$.  In this case the departure from the minimum of the
potential is small, $v_\mu^2 \simeq v^2$.
In this limit, the scalar field rolls around the circular minimum of
the potential, as illustrated in Figure \ref{fig:cartoon}.
The field possesses kinetic but almost no potential energy, and
the speed of sound approaches 1.  The addition of a thermal bath with
$T\gtrsim \mu$ will add a radiation component with
energy density $\varepsilon \simeq (N-1) \pi^2 T^4/30$, where $(N-1)$
counts the Goldstone bosons which will participate in the radiation.
Provided that $T^4 \ll \mu^2 v^2$, this is also a subdominant
contribution.

Why does a kinetic energy term of this form support
$P \simeq \varepsilon$?
Imagine changing the system's volume:
$\VV \to \VV+\delta \VV$ with $|\delta \VV| \ll \VV$.
The initial charge density was
$j^0_{12} = \mu v_\mu^2 \simeq \mu v^2$.  Charge conservation means that
$j^0$ must increase to keep $\VV j^0$ fixed:
$\delta j^0/j^0 = - \delta \VV/\VV$.
But the energy density is
$\varepsilon = \mu^2 v_\mu^2 \simeq \mu^2 v^2 \simeq (j^0)^2/v^2$.
Doubling the charge density required doubling the field velocity,
which quadruples the energy density.
Therefore $\delta \varepsilon/\varepsilon = -2 \delta \VV/\VV$.
In generality,
$\delta \varepsilon = -(\varepsilon+P) \delta \VV/\VV$, so this represents
$P=\varepsilon$.
In other words, because $\varepsilon \propto (j^0)^2$,
the energy density increases very rapidly under compression,
representing a very large pressure and supporting a speed of sound
approaching the speed of light.

To complete the calculation, expanding to include the first $\mu^4$
and $T^4$ corrections, the number density, energy density, and pressure
are given by
\begin{align}
  j^0 & \simeq \mu v^2 + \frac{2\mu^3}{\lambda} ,
  \nonumber \\
  \varepsilon & \simeq
  \half \mu^2 v^2 + \frac{3\mu^4}{2\lambda} + (N-1) \frac{\pi^2 T^4}{30},
  \nonumber \\
  P & \simeq
  \half \mu^2 v^2 + \frac{\mu^4}{2\lambda} + (N-1) \frac{\pi^2 T^4}{90}.
\end{align}
Under a volume change, $\delta T/T = -\delta \VV/3\VV$
and $\delta j^0 /j^0 = - \delta \VV/\VV$.
Therefore $\delta \mu/\mu = -(1-4\mu^2/\lambda v^2) \delta \VV/\VV$.
Inserting these shifts to find $dP/d\VV$ and $d\varepsilon/d\VV$,
we arrive at
\begin{equation}
  v_s^2 = \frac{dP}{d\varepsilon} = \frac{\VV\: dP}{d\VV}
  \left( \frac{\VV \:d\varepsilon}{d\VV} \right)^{-1}
  \simeq 1 - \frac{4 \mu^2}{\lambda v^2}
  - \frac{4(N-1)\pi^2 T^4}{135 \mu^2 v^2}
  \,.
\end{equation}

We see that scalar $O(N)$ $\lambda \varphi^4$ theory
has $v_s^2 \to 1$ when $T^2 \sim \mu^2 \ll m_r^2$.
This occurs because the energy density is dominated by a homogeneous
condensate which is maximally incompressible.

\section{Hydrodynamics of the model}
\label{sec:hydro}

Hydrodynamics is by definition the large distance and time behavior of
the conserved quantities in the system and their Noether currents.
In theories where only stress-energy is conserved, it is a theory of
the stress tensor $T^{\mu\nu}$, which typically deviates from its
equilibrium form at linear order as
\begin{align}
  \label{Thydro}
  T^{\mu\nu} & = \varepsilon \, u^\mu u^\nu + P \Delta^{\mu\nu}
  -\eta \left( \partial^\mu u^\nu + \partial^\nu u^\mu
  - \frac{1}{3} \Delta^{\mu\nu} \partial_\alpha u^\alpha \right)
  - \zeta \Delta^{\mu\nu} \partial_\alpha u^\alpha \,,
  \\ \nonumber
  \Delta^{\mu\nu} & \equiv g^{\mu\nu} + u^\mu u^\nu \,.
\end{align}
In our case there are $N(N-1)/2$ additional conserved Noether
currents, one of which carries a large value.
This complicates the general hydrodynamical equations,
which must be extended to include Goldstone excitations,
as discussed in Ref.~\cite{Son:2000ht,Pujol:2002na}.

Here we will concentrate on understanding the viscosity, without
worrying about the other transport coefficients which arise.
The viscosity and its relaxation time are defined in terms of a
Kubo formula, which reads \cite{Baier:2007ix}
\begin{equation}
  \label{Kubo}
  G^{xy,xy}_{R}(\omega,k) = P - i \eta \omega + \eta \tau_\pi \omega^2
  +\OO(k^2)
\end{equation}
where $G^{xy,xy}_R$ is the retarded correlator of two $T^{xy}$ stress
tensors in frequency $\omega$ and momentum $k$ space.
For our case the coupling is small and the interactions between light
modes are all derivative interactions, so a quasiparticle or kinetic
picture should be valid and we should be able to evaluate
$\eta,\tau_\pi$ using the kinetic theory approaches pioneered by
Baym et al \cite{Baym:1990uj} and by
Jeon and Yaffe \cite{Jeon:1994if,Jeon:1995zm} and extended to
second-order coefficients by Moore and York \cite{York:2008rr}.

First we need to expand $\epsilon - \mu j^0$ about the classical solution.
There are two approaches.  A common approach is to use angular and
radial variables.  In the absence of a chemical potential, the
potential then depends only on the radial variable, making the
masslessness and derivative coupling of the Goldstones manifest.  But
the kinetic term is nonlinear, giving rise to interactions.
This approach is somewhat less practical in our case because the
chemical potential further breaks the symmetry and not all angular
modes will turn out to be massless.
So instead we will work in terms of orthogonal field directions, but
we choose the fluctuations in the $(\varphi_1,\varphi_2)$ plane to
co-rotate with the condensate:
\begin{align}
  \varphi_1 & = (v_\mu + h) \cos(\mu t) + \pi \sin(\mu t) \,,
  \nonumber \\
  \varphi_2 & = (v_\mu + h) \sin(\mu t) - \pi \cos(\mu t) \,,
\end{align}
where $h,\pi$ represent the radial excitations and the angular
Goldstone mode respectively.
Inserting this into $\epsilon - \mu j^0_{12}$, and allowing the index $a'$ to run
over $3,\ldots N$, we find
\begin{align}
  \label{Hexpanded}
  \epsilon - \mu j^0_{12} = {}& \half \left( \partial_t \pi^2 + \partial_t
  h^2 + \partial_t \varphi_{a'}^2 \right)
  - \frac{\mu^2 v_\mu^2}{2\lambda}
  \nonumber \\
  & {} + \half \left( \partial_i \pi \partial_i \pi
  + \partial_i h \partial_i h
  + \partial_i \varphi_{a'} \partial_i \varphi_{a'}
  \right)
  \nonumber \\ 
  & {}   + \frac{\mu^2}{2} \varphi_{a'}^2 + \frac{\lambda v_\mu^2}{2} h^2
  + \frac{\lambda v_\mu}{2} h \left( h^2 + \pi^2 + \varphi_{a'}^2 \right)
  + \frac{\lambda}{8} \left( h^2 + \pi^2 + \varphi_{a'}^2 \right)^2 .
\end{align}
Several points should be made.
First, the $\varphi_{a'}$ field directions become massive with
$m^2 = \mu^2$, and are no longer Goldstone modes.
A fluctuation in one of these directions represents a local distortion
in the field direction in which the condensate varies, which will
naturally rotate at the same angular frequency $\mu$ as the condensate
itself.  However, $\pi$, representing a fluctuation in the phase of
the condensate, remains massless.
Second, the radial excitation $h$ has also increased in mass,
$m_r^2 = \lambda v^2 \to \lambda v_\mu^2$.
Finally, in this basis the Goldstone modes appear to possess
non-derivative cubic and quartic interactions.
We will now see how matrix elements nevertheless prove to represent
derivative couplings.

\begin{figure}[htb]
  \centerline{\includegraphics[width=0.8\textwidth]{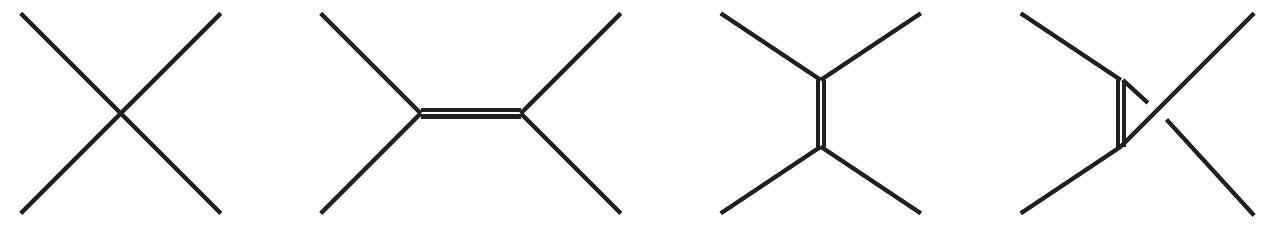}}
  \caption{\label{fig:fourpi} Feynman diagrams for the scattering
    $\pi \pi \to \pi \pi$.  The intermediate states are the radial
    mode $h$.}
\end{figure}

Consider first the case $N=2$.
Assuming $T,\mu \ll m_r$ the only constituents in the thermal bath are
the $\pi$ particles.  The lowest-order matrix element for
$\pi \pi \to \pi \pi$ scattering arises from the diagrams shown in
Figure \ref{fig:fourpi} and the Feynman rules yield
\begin{equation}
  \MM = 3 \lambda - (\lambda v_\mu)^2 \left(
  \frac{1}{m_r^2 + s} + \frac{1}{m_r^2 + t} + \frac{1}{m_r^2 + u}
  \right).
  \label{M1allpi}
\end{equation}
Here $(s,t,u)$ are the usual Mandelstam variables.
Recall that $m_r^2 = \lambda v_\mu^2$ which is much larger than
$s,|t|,|u| \sim T^2$.  Performing a geometric series expansion, we find:
\begin{equation}
  \label{M2allpi}
  \MM = 3 \lambda - 3 \lambda
  + \frac{1}{v_\mu^{2}} \left( s + t + u \right)
  - \frac{1}{v_\mu^4 \lambda} \left( s^2 + t^2 + u^2 \right) + \ldots
\end{equation}
The leading terms cancel, which is how the derivative-coupled nature
of Goldstone boson interactions reasserts itself in our field basis.
But $s+t+u=0$ in massless kinematics, so the second term also cancels.
The resulting matrix element is $\propto s^2$ and the squared matrix
element scales as $\propto s^4$.
When one of the particles in a scattering process has a small momentum
$p \ll T$, we have $s^4 \propto p^4$.
Soft modes therefore have a lifetime in the plasma proportional to
$p^{-4}$.  The standard calculations of transport coefficients
\cite{Arnold:2003zc} then predict that the departure from equilibrium
in the presence of shear flow, represented by $\chi(p)$ in the
notation of that reference, will scale as
$\chi(p) \propto p^{-3}$ and the shear viscosity will behave as
\begin{equation}
  \label{etadiverge}
  \eta \propto \int p^2 \, dp \; p n_b(p) [1{+}n_b(p)] \chi(p)
  \sim \int dp\; p^3 \frac{T^2}{p^2} p^{-3}
\end{equation}
which is small-$p$ divergent.  Therefore the theory for $N=2$ does not
have a well behaved shear viscosity if we only consider Goldstone
bosons.  Damping of the most IR Goldstone bosons will be dominated by
scattering off the exponentially rare radial modes, leading to an
exponentially large shear viscosity.

\begin{figure}[htb]
  \centerline{\includegraphics[width=0.5\textwidth]{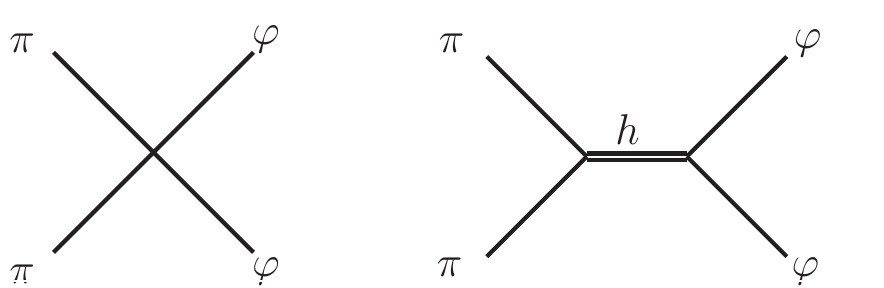}}
  \caption{\label{fig:piphi} The two Feynman diagrams for the scattering
    $\pi \pi \to \varphi \varphi$.}
\end{figure}

The case $N > 2$ is better behaved.  Consider $N=3$ for
concreteness.  There is now an additional light mode $\varphi_3$
which I will simply write $\varphi$.  The scattering matrix element
for $\pi \pi \to \varphi \varphi$, arising from the diagrams of Figure
\ref{fig:piphi}, is
\begin{equation}
  \label{M1piphi}
  \MM = \lambda - \lambda^2 v_\mu^2 \frac{1}{m_r^2 + s}
  \simeq \frac{s}{v_\mu^2} \,.
\end{equation}
For $\pi \varphi \to \pi \varphi$ scattering we have
$\MM = t/v_\mu^2$ by crossing.  If we had used angular and radial
variables for our fields, the relevant interaction would arise from
the curvature of the vacuum-manifold sphere, which explains why the
interaction is suppressed by two powers of the radius of the vacuum
manifold $v_\mu$.
The form of the matrix element is the same as for scattering between
different pion species in the small $m_\pi$ limit \cite{Scherer:2002tk}
and was originally derived by Weinberg \cite{Weinberg:1966kf}.
Assuming $T \sim \mu$ so that the $\varphi$ particles are present in the
medium, the scattering leads to a damping of long-wavelength $\pi$ modes
with decay rate $\Gamma \propto p^2$, a departure from equilibrium
$\chi(p) \propto p^{-1}$, and a convergent value for the viscosity;
\Eq{etadiverge} has two fewer inverse powers of $p$ and therefore
converges at small $p$.

We have carried out a concrete calculation for the case $N=3$ and
treating $T \gg \mu$ so that both fields obey massless kinematics.
Using the methodology of \cite{Jeon:1995zm,Arnold:2003zc},
we find explicitly that
$\eta = 3.38083 v_\mu^4/T$.  $\eta/s \sim (v/T)^4$ is parametrically
large, but this is expected given that Goldstone modes are weakly
coupled.

Unfortunately the shear-viscous relaxation rate is not well behaved.
According to Ref.~\cite{York:2008rr}, this is
\begin{equation}
  \eta \tau_\pi \propto \int p^2 \, dp \; n_b(p)[1{+}n_b(p)]
  \chi^2(p)
\end{equation}
which given our previous result, $\chi(p) \propto p^{-1}$,
and $n_b(p) \simeq T/p$ for the massless Goldstone modes,
leads to a linearly IR divergent contribution.

Let us explore the physical reason that $\tau_\pi$ diverges.
Our theory contains Goldstone modes which propagate as hydrodynamical
modes with decay rates scaling as $\Gamma(p) \propto p^2$ for small
$p$.  Such hydrodynamical modes give rise to so-called ``long-time
tails'' in correlation functions of stress tensors
\cite{Pomeau:1975,Kovtun:2003vj},
such as the one which defines $\eta$ and $\eta \tau_\pi$.
It was shown by Kovtun, Moore, and Romatschke \cite{Kovtun:2011np}
that sound waves, which also have $\propto p^2$ decay rates, also
give rise to a contribution to \Eq{Kubo} which is nonanalytic,
$G^{xy,xy}_R(\omega) = P -i\eta \omega + \OO(\omega^{3/2})$.
Strictly speaking, such a term renders $\eta \tau_\pi$ ill defined, or
alternatively, infinite.
There are cases where the coefficient in front of $\omega^{3/2}$ is small
enough that it is dominated by both the $\omega$ and an $\omega^2$
term over some range of frequencies.  In such a case, there is an
intermediate range of scales where second-order hydrodynamics can be
applied.

In the absence of Goldstone modes, where sound modes are the principal
culprit in such long-time tails, such a regime can exist in either of
two cases. If the shear viscosity is large $\eta/s \gg 1$, then the
sound waves are rather efficiently damped except for very small $p$.
Hence the long-time tails only emerge at very small $\omega$.
Also, if the number of degrees of freedom is large, then the
contribution to $G^{xy,xy}_R$ from sound modes is overwhelmed by the
contributions from the large number of degrees of freedom, except
again at very small $\omega$.
Both of these possibilities are discussed in Ref.~\cite{Kovtun:2011np}.

But in our case, the main problem arises from the long-time tails
arising from the Goldstone modes.  Such modes are unavoidable in this
model because we need a scalar condensate to provide the large pressure.
Weaker coupling leads to a more weakly-damped Goldstone mode and
therefore a larger $\omega^{3/2}$ coefficient.
And the limit of $N \to \infty$, to increase the degree-of-freedom
count, risks providing large additional contributions to the pressure
and energy density which do not obey $P \simeq \varepsilon$.

In conclusion, we find that the shear viscosity is finite and
calculable for $N>2$, but that the long-time tails arising from
Goldstone modes' contribution to the viscosity make it impossible to
define $\tau_\pi$, or rather, the contributions of such modes to
$\tau_\pi$ diverges.  This explains why this model can violate the
proposed bound of Hippert, Noronha, and Romatschke.

\section{Discussion}
\label{sec:discussion}

We see that finding theories with $v_s^2 \to 1$ is rather easy if one
invokes a chemical potential.
However, such chemical potentials together with scalar condensates
lead to complicated hydrodynamics, and the presence of Goldstone modes
generally leads to a divergently large shear relaxation time, due to
the ``long-time tails'' which arise from the contribution of
slowly-equilibrating infrared Goldstone modes to the shear viscosity.
As emphasized in the last section, such long-time tails actually
always occur due to sound modes and $\tau_\pi$ is never strictly well
defined, as discussed by Kovtun Moore and Romatschke
\cite{Kovtun:2011np}.
However, in many cases there is a broad range of scales where
$\tau_\pi$ is approximately scale invariant; but for the theories
considered here this will \textsl{not} be the case, since the relevant
Goldstone modes contribute an $\OO(1)$ fraction of the shear viscosity.

One might object that the theories considered are not asymptotically
free and suffer from a Landau pole.
But there is at least one case where we know a UV completion.
$SU(3)$ gauge theory with two massless fundamental vectorlike fermion
fields (quarks) exhibits the spontaneous breaking of an
$SU_L(2)\times SU_R(2)$ flavor symmetry to $SU_V(2)$,
with an $SU(2) \cong S^3$ vacuum manifold.
The low-energy effective theory is like our $N=4$ scalar theory after
integrating out the radial mode, and Son and Stephanov have already
shown \cite{Son:2000by} that it features a speed of sound approaching
the speed of light.  Hopefully this paper will then help to explain
the simplicity of the physics which gives rise to $v_s^2 \to 1$ in
this case.

We emphasize that the role of the chemical potential is central to
our example, and indeed, all examples we know where $v_s^2$ really
significantly exceeds $1/3$.
But there are nevertheless examples of theories with
$v_s^2 > 1/3$ without the need for chemical potentials, provided that
one does not mind a Landau pole at scales orders of magnitude larger
than the thermal scale.
Massless QED and massless scalar $\lambda \varphi^4$ theory both fall
in this category.  To see this, note that the pressure can generically
be expanded in the temperature and the coupling $g$ as
\begin{equation}
  P(g,T) = (A - B g^2 + \ldots) T^4
\end{equation}
with $A,B$ some theory dependent constants with $A>0$ and $B>0$ in
most cases, see for instance
\cite{Kapusta:1979fh,Arnold:1994eb,Braaten:1995jr,Kajantie:2002wa}.
Calling the beta function
\begin{equation}
  \frac{\mu^2\, d\,g^2}{d\mu^2} \equiv \beta_0 g^4
\end{equation}
the renormalization point independence of the pressure and the fact
that $T$ is the only physical scale ensures that
\begin{equation}
  P(g,T) = \left( A - B g^2 + B \beta_0 g^4 \ln \frac{\mu^2}{T^2}
  + \ldots \right) T^4
\end{equation}
where $\ldots$ has explicit $\mu,T$ dependence only at higher order
than what is shown.
Taking the temperature derivative, one finds that
\begin{align}
  \varepsilon & = T \frac{dP}{dT} - P
  = \left( 3A - 3B g^2 + 3 B \beta_0 g^4 \ln \frac{\mu^2}{T^2}
  - 2 B \beta_0 g^4 \right) T^4
  \\
  \frac{dP}{d\varepsilon} & \simeq \frac{1}{3}
  + \frac{2 B \beta_0 g^4}{9 A}
\end{align}
which is positive for $\beta_0 > 0$, as occurs for scalar field theory
and QED.  Therefore, in these non-asymptotically-free theories, the
speed of sound of an ordinary thermal medium without net conserved
charge densities exceeds $1/\sqrt{3}$, albeit by a perturbatively
small amount.
Note that this simple example features theories which are not UV
complete.  It would be interesting if one could find a UV completion
of such a theory, which would represent a rigorous example of a theory
with $v_s^2 > 1/3$ without chemical potentials playing any role.

\section*{Acknowledgments}

I would like to thank Paul Romatschke for convincing me to write this
up, and Gergeley Endr\H{o}di, Nicolas Wink, and Isabella Danhoni for
helpful conversations.
I acknowledge support by the Deutsche Forschungsgemeinschaft (DFG,
German Research Foundation) through the CRC-TR 211 ``Strong-interaction
matter under extreme conditions''– project number 315477589 – TRR 211
and by the State of Hesse within the Research Cluster ELEMENTS
(Project ID 500/10.006).

\bibliographystyle{unsrt}
\bibliography{refs}

\end{document}